\documentclass{mem}
\usepackage{natbib}\usepackage{txfonts}\usepackage{balance}
\usepackage{graphicx}
\usepackage[a4paper]{hyperref}
\idline{75}{282}
\begin{document}

\title{
latitude of Ephemeral Regions as Indicator of Strength of  
Solar Cycles
}
   \subtitle{}

\author{
Andrey G. \, Tlatov\inst{1} 
\and Alexei A. \, Pevtsov\inst{2}
          }

  \offprints{A.G.  Tlatov}

\institute{
Kislovodsk Solar Station of Pulkovo Observatory, Russian Federation\\
\email{tlatov@mail.ru}
\and
National Solar Observatory, Sunspot, NM 88349, U.S.A.
\email{apevtsov@nso.edu}
}

\authorrunning{Tlatov and Pevtsov}

\titlerunning{Latitude of Ephemeral Regions and Strength of Solar Cycles}

\abstract{
Digitized images of full disk CaK spectroheliograms from two solar 
observatories were used to study cycle variation of ephemeral regions (ERs) 
over ten solar cycles 14-23. 
We calculate monthly averaged unsigned
latitude of ERs and compare it with annual sunspot number.
We find that average latitude of ERs can be used as a predictor for strength 
of solar cycle. For a short-term prediction (dT $\sim$ 1-2 years), maximum 
latitude of 
ephemeral regions (in current cycle) defines the amplitude of
that cycle (higher is the latitude of ERs, larger are the amplitudes of 
sunspot cycle).
For a long-term prediction (dT $\sim$ 1.5 solar cycles), latitude of 
ERs at declining phase of n$^{th}$ cycle determines the amplitude of 
(n+2)$^{th}$ sunspot cycle (lower is the latitude of ERs, stronger is the 
cycle). Using this latter dependency, we forecast the amplitude of sunspot 
cycle 24 at W=92$\pm$13 (in units of annual sunspot number).

}
\maketitle{}

\section{Introduction}
\begin{figure}[]
\resizebox{\columnwidth}{!}{\includegraphics[clip=true]{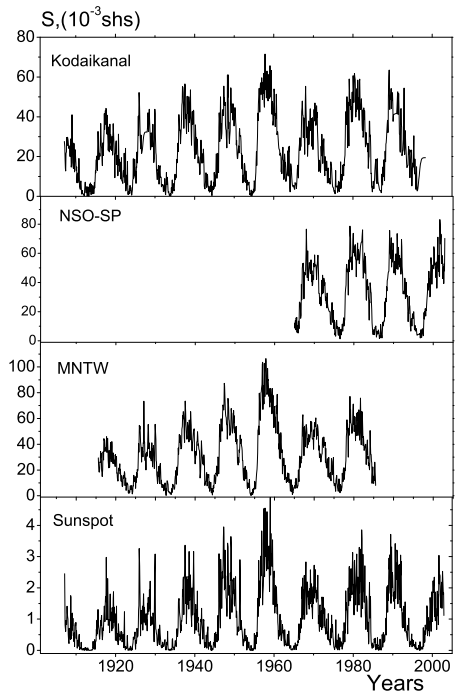}}
\caption{\footnotesize
Monthly averaged area of Ca II K plages derived from data from three 
observatories (three upper panels) and area of sunspots (lower panel).}
\label{fig:ca_ex}
\end{figure}

Historic data sets of full disk Ca II K spectroheliograms observed from three 
observatories: Kodaikanal (KKL), Mount Wilson (MWO), and the National 
Solar Observatory at Sacramento Peak (NSO/SP) span about ten past solar cycles. 
Recently, these data were digitized and calibrated. In this study we use KKL and 
NSO/SP datasets to explore latitudinal distribution of ephemeral active regions 
as potential precursor of amplitude of sunspot cycles.
Images were acquired with exit slit of spectroheliograph of about 0.5 \AA\ in 
width centered at $\lambda$ = 3933.67\AA. Spatial resolution is 
approximately 1.2 
arc seconds per pixel. Further details on these data sets can 
be found in \cite{Tlatov2009}. Flocculi and plages were 
identified using intensity threshold as described in \cite{Tlatov2009}. 
Isolated clusters of bright pixels (not connected with other clusters) were 
classified as independent flux elements, and the total number of elements and 
area of each element were computed for every image in our data set. Despite 
differences in observations, floccular areas show good correlation between 
instruments and with sunspot number (Figure \ref{fig:ca_ex}).
\cite{Tlatov2009} established a correlation between total area of CaK 
plages ($A_{plage}$) and area of sunspots ($A_{sunspot}$),
$A_{plage}=(8.5\pm0.3)+(15\pm0.25)A_{sunspot}$, with Pearson linear correlation 
coefficient R=0.88.
KKL data set is the longest of three, while NSO/SP set is the most recent.
Given good correlation between 
KKL, MWO, NSO/SP Ca II K data we chose to use the Kodaikanal and NSO/SP 
data for the following analysis.

\section{Average Latitude of Ca II K Flocculi}

\begin{figure}[]
\resizebox{\columnwidth}{!}{\includegraphics[clip=true]{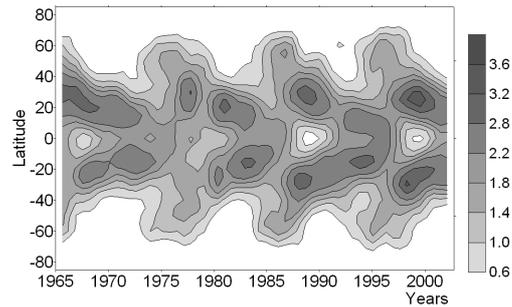}}
\caption{\footnotesize
Number of Ca II K ephemeral regions in NSO/SP data as function of time and 
latitude. Data corresponds to 3-months averages computed over 5$^\circ $ latitudinal intervals.  
}
\label{fig:sp_lat}
\end{figure}

We have selected a subset of mid-size floccular elements with area between 100 
and 300 millionths of solar hemisphere (MSH). \cite{Harvey1973} had 
classified elements with this area observed in Ca II K line as ephemeral 
regions. Majority of bright elements identified in Ca II K images correspond to 
the chromospheric network elements; only a small fraction ($\sim$ 10\%) matches 
ephemeral regions and plages. Furthermore, while total number of bright 
elements does not vary with sunspot cycle, number of ephemeral regions does 
exhibit sunspot cycle-like variations. 
Figure \ref{fig:sp_lat} shows number of ERs found in NSO/SP data as function of 
time and latitude. Typically, cycles of ephemeral regions begin at latitudes of
$\sim$ 60 degrees about one-two years before sunspot maximum.
ERs cycles appear to be shifted relative to 
sunspot cycles: number of ERs is at its lowest at the rising phase of sunspot 
cycle, and it reaches maximum in the tail of sunspot cycle when sunspot 
activity concentrates at the equatorial region.  

Next, we have calculated average latitude of ephemeral regions $\bar\theta = {1 
\over N} \displaystyle\sum_{i=1}^{N}\vert\theta_i\vert,$ where N is total number of ERs 
observed in a given month and $\theta_i$ is the latitude of individual ER.

\begin{figure}[]
\resizebox{\columnwidth}{!}{\includegraphics[clip=true]{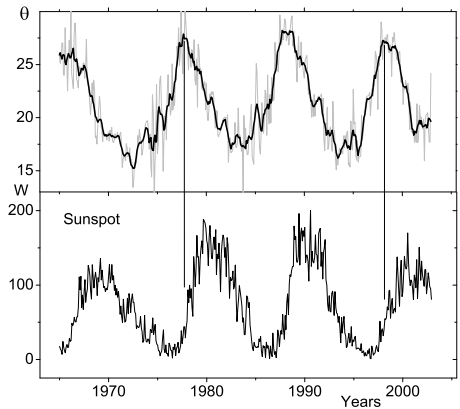}}
\caption{\footnotesize
Three months (upper panel, thin curve) and 12-months (thick curve) averaged 
latitude of 
Ca II K ephemeral regions derived from NSO/SP data and annual sunspot number 
(lower 
panel). Thin vertical lines demonstrate that ephemeral regions reach maximum 
latitude 
at rising phase of sunspot cycle, a few years prior to sunspot maximum.}
\label{fig:sp_cycle}
\end{figure}

\begin{figure}[]
\resizebox{\hsize}{!}{\includegraphics[clip=true]{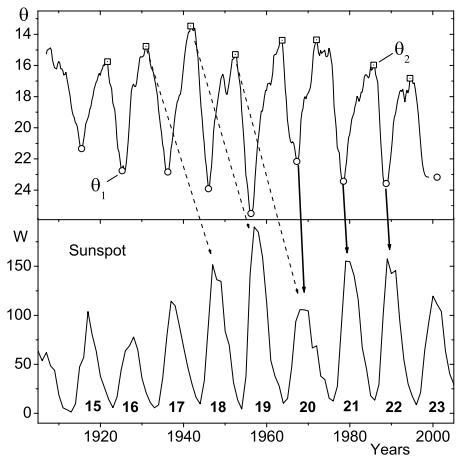}}
\caption{\footnotesize
Average latitude ($\bar\theta$, in degrees) of ephemeral regions (upper panel) and 
sunspot 
number (W, lower panel). Open squares and circles mark local minimum ($\bar\theta_{2i}$) 
and maximum 
($\bar\theta_{1j}$) latitudes of ERs. Dashed lines with arrowheads draw a 
correspondence 
between minimum latitude of ERs and maximum sunspot number. Solid lines with 
arrowheads show correspondence between maximum latitude of ERs and maximum 
sunspot number. To see the pattern, compare order of cycles in their amplitude 
with order of minima and maxima in average latitude of ERs.}
\label{fig:er_cycle}
\end{figure}

\begin{figure}[]
\resizebox{\hsize}{!}{\includegraphics[clip=true]{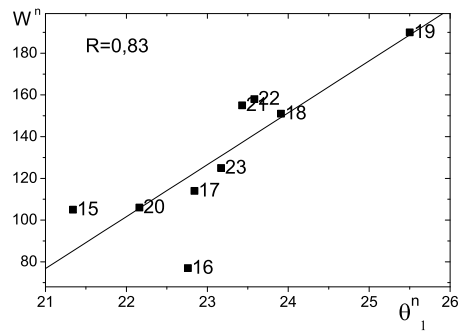}}
\caption{\footnotesize
Amplitude of n$^{th}$ sunspot cycle (annual sunspot number) as function of maximum 
latitude (in degrees) of ERs in rising 
phase of same cycle. Solid line shows a linear approximation to the data,
$W^{n}=(-445.782 \pm 146.000) + (24.884 \pm 6.289) \times \bar{\theta}^{n}_1$.}
\label{fig:c1_plot}
\end{figure}

Average latitude of ephemeral regions shows clear variation with sunspot cycle 
(Figure \ref{fig:sp_cycle}). ERs latitudes reach maxima on the rising phase of 
sunspot cycle, approximately one -- two years before the number of 
sunspots reaches its maximum. Examining Figures \ref{fig:sp_cycle} and 
\ref{fig:er_cycle} one may note a correlation between maximum latitude 
$\bar\theta_1$ of ephemeral regions and amplitude 
of sunspot cycle: higher is the average latitude of ERs, stronger 
is the sunspot cycle. Similar tendency is also present in the butterfly 
diagram of sunspots (not shown): in stronger cycles first sunspots emerge at 
higher 
latitudes, while for weaker cycles, latitude at which early sunspots emerge 
is lower. 
Figure \ref{fig:c1_plot} shows correlation between maximum latitude of ERs 
($\bar\theta_1$) and sunspot number (W); Pearson linear correlation 
coefficient 
R$_{\theta 1}$ = 0.83. 

Even more interesting, minima of average latitude of ERs ($\bar\theta_2$) 
that occur on declining phase of n$^{th}$ cycle correlate quite well with 
amplitude of (n+2)$^{th}$ sunspot cycle. Dashed lines with arrowheads shown in 
Figure \ref{fig:er_cycle} demonstrate this relation. 

\begin{figure}[]
\resizebox{\columnwidth}{!}{\includegraphics[clip=true]{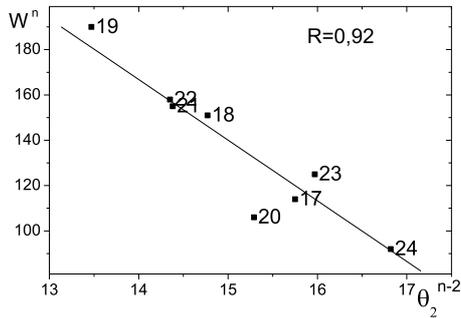}}
\caption{\footnotesize
Amplitude of n$^{th}$ sunspot cycle (annual sunspot number) as function 
of minimum latitude (in degrees) of ERs 
in (n-2)$^{th}$ cycle. Solid line shows linear approximation to the data
$W^n=(541.467 \pm 69.939) - (26.761 \pm 4.622) \times \bar\theta^{n-2}_2$.}
\label{fig:c2_plot}
\end{figure}

Figure \ref{fig:c2_plot} shows correlation between minimum latitude of ERs 
($\bar\theta_2$) and sunspot number (W) with R$_{\theta 2}$=0.92. This latter 
correlation enables forecast of amplitude of sunspot cycle for about one and a 
half solar cycles prior to its occurrence.

\section{Discussion}

Forecasting the strength of future solar cycles was attempted by number of 
researchers \citep[e.g.,][]{Ohl1968,Javaraiah2007}. In our present study, we 
demonstrate that average latitude of ephemeral active regions can be used for a short-term (dT$_1$ $\sim$ 
1--2 years) and a long-term (dT$_2$ $\sim$ 1.5 solar cycles, or 14--17 years) 
forecast of amplitude of sunspot cycle. Appearance of emerging regions at high 
latitudes 
heralding emergence of sunspots is in agreement with the idea of extended solar 
cycle \citep{Wilson1988} with magnetic activity of a current cycle starting at 
high latitudes a few years prior to sunspot emergence. In its turn, dT$_2$ is 
very close to a time interval to transport surface magnetic field into the 
dynamo region as required by the transport dynamo models \citep{Choudhuri1995,Tlatov1996}. Latitude $\bar\theta_2$ is lower, when the old cycle activity 
penetrates to lower latitudes in absence of high-latitude activity of a new 
cycle. This implies that $\bar\theta_2$ is lower, when overlap between old and new 
cycles is shorter. Since the orientation of magnetic fields in two consecutive 
cycles (cycles n and n+1) is opposite to each other, longer overlap between 
cycles will lead to a weakening of a seed field for (n+2) cycle. In 
addition, a penetration of magnetic field to lower latitudes may allow for a 
more efficient cancellation of magnetic field of opposite polarity across solar 
equator, which in the framework of Babcock-Leighton model may lead to a more 
effective transformation of magnetic fields of active regions into poloidal 
field of future cycles.

Finally, using found relationship between minimum latitude of ephemeral regions 
and amplitude of solar cycle (Figure \ref{fig:c2_plot}), we predict a relatively 
modest cycle 24 with annual averaged amplitude of about W=92$\pm$13. This our 
prediction is in agreement with recent forecast of strength of Cycle 24 
(92.8$\pm$19.6) by \cite{Bhatt2009} based on geomagnetic aa index.   

\begin{acknowledgements}

A.T. work was supported by the Russian Foundation for Basic Research (RFBR), the 
Russian Academy of Sciences, and the Program to Support Leading Scientific 
Schools by the Russian Federal Agency for Science and Innovations. 
A.P. acknowledges partial support from NASA's NNH09AL04I interagency transfer. 
The National Solar Observatory is operated by the Association for Research in 
Astronomy (AURA, Inc.) under cooperative agreement with the National Science 
Foundation (NSF).

\end{acknowledgements}

\bibliographystyle{aa}

\end{document}